\documentclass{article}

\usepackage{mathptmx}

\usepackage{natbib}
\bibpunct{[}{]}{,}{n}{}{,}

\begin{document}

\title{Quantum Measurement, Complexity, And Discrete Physics}
\author{Martin J. Leckey\\School of Historical and Philosophical Studies\\The University of Melbourne, Victoria 3010, Australia\\mjleckey@unimelb.edu.au}
\date{21 April 2016}

\maketitle

\newcommand{\x}{{\rm\bf x}}

\begin{abstract}
This paper presents a new modified quantum mechanics, Critical Complexity Quantum Mechanics,  which includes a new account of wavefunction collapse.  This modified quantum mechanics is shown to arise naturally from a fully discrete physics, where all physical quantities are discrete rather than continuous.  I compare this theory with the spontaneous collapse theories of Ghirardi, Rimini, Weber and Pearle and discuss some implications of the theory for a realist view of the quantum realm.
\end{abstract}

\section{Introduction}
\label{Sec:Intro}
In this paper, a proposal for a modified quantum mechanics is outlined. This modified quantum mechanic includes an outline of a solution to the ``measurement problem'' of quantum mechanics. However, the modifications to quantum mechanics introduced are not motivated merely by a desire to solve the measurement problem. I contend that the theory can be shown to arise naturally from a fully discrete physics, where all physical quantities are discrete rather than continuous. Briefly put, the idea is that because wavefunctions always expand or spread with time, eventually they will grow to such an extent that the average magnitude of the wavefunction threatens to fall below the minimum magnitude representable in the discrete physics under consideration, thus threatening to disappear altogether.  I argue that in order to prevent this happening, it is natural---that is, theoretically well motivated---that the wavefunction will ``collapse'', that it will suddenly reduce in size, when it reaches a certain critical ``volume'' in configuration space. This not only brings about a solution to the measurement problem but also motivates the idea that there are multiple wavefunctions in the universe, not just a single wavefunction for the whole universe. The result is a theory which allows for a realist interpretation of the wavefunction. While this paper is based on a presumption of a discrete physics, I suggest that a version of the theory can stand independently of whether physics is continuous or discrete. 

After providing a brief account of the measurement problem, I introduce the new theory, which I call Critical Complexity Quantum Mechanics (CCQM). I then provide an account of discrete physics as applied to quantum mechanics and set up the motivation for the new theory. After presenting more details of CCQM, I outline how it may be able to provide a solution to the measurement problem. The theory has some features in common with two closely related theories of wavefunction collapse that have been proposed with the intention of providing a solution to the so-called ``measurement problem'' of quantum mechanics. These are the theories of Ghirardi, Rimini and Weber \citep{grw1986, grw1988} and the continuous spontaneous localisation (CSL) theory \citep{gpr1990}. I give a short account of the Ghirardi, Rimini and Weber (GRW) theory for the purpose of comparison. Finally I give some indication of what some details of the wavefunction collapse model might look like, and compare the new theory with the GRW and CSL theories. Finally, I discuss the realist interpretation of the wavefunction that the theory allows. 

\section{The Measurement Problem}
\label{Sec:MeasProb}
The measurement problem arises when we consider applying the standard linear equation of motion of quantum mechanics, the Schr\"{o}dinger equation, to macroscopic systems. The problem is seen acutely when one tries to describe measurement processes; that is, when one treats the measuring apparatus not as separate from the measured system, but as part of a larger system, and describes the evolution of this larger system by the Schr\"{o}dinger equation.  This evolution does not give rise to a determinate outcome for the measurement, but instead a superposition of different possible outcomes---a superposition of macroscopically distinguishable states (assuming that an observable does not have a definite value for a state unless that state is an eigenstate of that observable.)  These macroscopic superpositions are never observed, and the problem of reconciling what we observe with the predictions of standard linear quantum mechanics is what I take to be the measurement problem.

The orthodox (Dirac-von Neumann) way to avoid this measurement problem is to assume the normal evolution of the state is suspended, so that the quantum system will change indeterministically when a ``measurement'' takes place, abruptly ``collapsing'' so as to give rise to a determinate measurement outcome.  However  no satisfactory account is given in orthodox quantum mechanics of what a ``measurement'' is, or how it brings about the collapse.  This situation is unacceptable if we assume that measurements are physical interactions and we seek a fundamental theory that can provide an account of these physical interactions without reference to an imprecise concept such as ``measurement''.   

Accordingly, many have attempted to introduce modifications to the linear evolution of the wavefunction, modifications that will eliminate the macroscopic superpositions that are not observed, via the collapse of the wavefunction, but leaving the quantum mechanics of microscopic systems largely unaltered, so that there will be no conflict with experimental results.  In this way it is hoped that references to measurements or observers can be eliminated from the fundamental formulation of the theory, and a unified dynamics can be provided for microscopic and macroscopic systems.  It is these kinds of responses to the measurement problem that will be considered in this paper.

\section{Critical Complexity Quantum Mechanics}
\label{Sec:CCQM}
I here introduce a modified quantum mechanics that not only provides a new approach to solving the so-called ``measurement problem'' but admits a realist interpretation of wavefunctions of lesser extent than a wavefunction representing the entire universe, and a mechanism for wavefunction collapse. A strength of the theory is that it can be motivated by what follows from considering a fully discrete physics. I will show that the collapse of the wavefunction is well motivated when considering a discrete physics, and is not a mere ad hoc modification to linear quantum mechanics. A brief outline of the theory will be given before discussing discrete physics.

According to standard quantum mechanics, as long as there is non-zero strength of interaction between particles they must be represented by a single configuration space wave function, whose arguments are the positions of all the particles.  Since there are presumably no truly isolated systems, it would seem that quantum mechanics leads to the necessity of representing all the particles of the universe in a single wavefunction. However, under many interpretations of quantum mechanics, wavefunctions are regarded as merely convenient constructions useful for making calculations. Under this anti-realist interpretation of wavefunctions, under no circumstances will the wavefunction be regarded as corresponding to what exists in nature.

The theory that I put forward here differs considerably from standard quantum mechanics.  It differs in providing for the existence of wavefunctions that represent a finite number of particles, much smaller than the total number of particles in the universe, and in allowing a realist interpretation of these wavefunctions and of wavefunction collapse.  Furthermore under this theory the number of particles a wavefunction represents can change with time, either through the wavefunction splitting into two or more smaller wavefunctions, or by combining with one or more other wavefunctions to form a larger one.  Although a wavefunction may vary in the number of particles it represents with time, this number will always be considerably less than the total number of particles in the universe.  From this point of view it is perhaps natural that a collapse or split of a wavefunction be triggered when it reaches some critical size, or complexity, for some measure of the complexity of the wavefunction. This takes up a suggestion by \citet[p.598]{leggett1984} that there may be ``corrections to linear quantum mechanics which are functions, in some sense or other, of the \textit{degree of complexity} of the physical system described''.

I propose the measure of complexity of a wavefunction (for the moment neglecting spin) is given by its ``relative volume'' in configuration space.   Configuration space has $3N$ dimensions for a system of $N$ particles so that the dimensionality of a ``volume'' in this space will vary with the number of particles represented by the wavefunction.  A quantity of consistent dimensionality is required in order to be able to compare the complexity of wavefunctions.  This requirement can be met by means of dividing the configuration space up into cells of small volume.  This enables a \textit{dimensionless} quantity for each wavefunction to be defined; namely, the number of these cells contained within the boundary of the wavefunction in configuration space.  This I define as the ``relative volume'' of that wavefunction.  I will assume that the wavefunction of a quantum system will collapse, or split, when it reaches a certain critical relative volume in configuration space.  What a ``collapse or split'' of the wavefunction amounts to will be explained in due course. The modified quantum mechanics that involves an altered quantum dynamics associated with this critical complexity of wavefunctions I label Critical Complexity Quantum Mechanics (CCQM). 

The small reference cells in configuration space can be defined by the introduction of a small reference length, or reference lengths.  (The reference lengths could be the same for all particles or vary with the particle and its energy.)  These reference lengths form the sides of the reference cell in configuration space. A definition of what is meant by the boundary of the wavefunction is also required. Consider for example a single particle, for which configuration space is ordinary three-dimensional position space.  In standard quantum mechanics the wavefunction of a single particle is taken to vary continuously over all position space, and so never drops to zero magnitude - if this were strictly true, then the volume occupied by the single-particle wavefunction would be the volume of the universe. This provides a problem for a theory that wishes to define a finite volume for the wavefunction. This problem that exists for the case of a continuous wavefunction can be overcome by means of a small reference magnitude $f_{0}$.  The measure of the wavefunction volume is taken to be given by the sum of the volumes of all the reference cells occupied by the wavefunction.  A reference cell is said to be occupied when the average wavefunction magnitude inside a reference cell is greater than $f_{0}$. 

This definition of the relative volume of a continuous wavefunction divides configuration space into discrete cells, and disregards that part of the wavefunction having magnitude less than a certain discrete magnitude. If it were the case that the wavefunction were in reality discrete, able to take on only a discrete range of values, and taking on only one value per cell in a cellular configuration space, then there would be some independent motivation for the modified quantum mechanics presented here. A fuller account of the motivation for CCQM provided by discrete physics is developed in the following sections, starting with a description of discrete physics.

\section{Discrete Physics}
\label{Sec:DisPhys}
Discrete Physics is characterised by the quantities representing the state of a system being discrete valued and finite in number \citep{feynman1982, fredkin1990, minsky1982, zuse2013}.\footnote{A collection of articles about physics as computation, published in 2013, contains a number of articles that posit a discrete physics \citep{zenil2013}.} This is the approach taken by those modelling physics using cellular automata \citep{wolfram1986, wolfram2002}.  In a cellular automaton the state of a physical system is taken to be represented by a certain finite number of discrete magnitudes, and these quantities are taken to be defined only at points on a spatial lattice, whereby the points of the lattice are separated by some finite distance.  These magnitudes are updated at a certain discrete time interval according to some rules, representing the fundamental laws of physics.  This approach is normally taken as a means of modelling continuous physical phenomena, but if it turned out that all state-variables currently taken to be continuous were in fact discrete, then an approach like this could give a more realistic picture of nature than an approach based on continuous quantities. 

Aspects of string theory and other quantum theories of gravity have led many authors to suggest that there may be a minimum length scale in the universe, at approximately the Planck length: $10^{-33}$ cm. (For example, see \citet{Witten1996}.) To many, such as \citet[p.152]{Hooft1997}, this suggests that space-time may be discrete rather than continuous. It is a further assumption that all physical quantities are discrete rather than just space and time, but 't Hooft \citep[pp.176--177]{hooft1993dimensional, hooft2001holographic, Hooft1997} does refer to Fredkin's ideas on cellular automata as indicating a possible way forward for physics. More recently, \citet{beane2014} have considered possible observational consequences of a discretized space-time.

I will adopt in this paper the following heuristic principle: consider how a discrete physics could model physical laws, then only accept as possible laws of nature those physical laws that could be successfully modelled this way.  The heuristic principle has clear justification if nature is in reality discrete, but it can also be used as a means of picking out laws that apply to continuous physics, since there are usually laws of continuous physics that correspond closely to the laws expressed in terms of discrete quantities.  I claim that it is an advantage of the critical complexity theory that it seems to follow naturally from the discretization of quantum mechanics.  Accordingly I will define a discrete-valued wavefunction corresponding to the continuous-valued one, and this will be taken as my starting point.  However I propose that a version of the theory can stand independently of whether nature is in fact continuous or discrete.  In the case where the wavefunction is taken to be continuous, the discrete wavefunction can be seen as having merely heuristic value.  In that case the base magnitude $f_{0}$ and the lengths defining the cell size in configuration space are taken as ``purely reference'' magnitudes - magnitudes in nature that play a role in defining a finite volume for a continuous wavefunction.  In the case of a physical world that is in reality discrete,  $f_{0}$ and the reference lengths may take on further significance; namely, as magnitudes that characterize truly discrete quantities in nature. 

\section{The Discrete Physics Representation}
\label{Sec:DisPhysRep}
In this section I will describe one way quantum wavefunctions could be represented in terms of a discrete physics.  The discretization of the wavefunction will be carried out in the position  representation.  In the following section the possible consequences of this discreteness will be investigated.

A many-particle system of $N$ particles can be described by a wavefunction in configuration space.  Consider the wavefunction for a system of $N$ interacting particles, neglecting spin.  In standard quantum mechanics, the wavefunction is continuous in magnitude and defined at all points in configuration space.  Write the wavefunction as a product of a magnitude and phase factor as follows:
\begin{equation}
\psi (\x_{1} ,\ldots ,\x_{N} ,t) = f(\x_{1} ,\ldots ,\x_{N} ,t)e^{i\theta (\x_{1} ,\ldots ,\x_{N} ,t)}.
\end{equation}
The real-valued function $f$ is the magnitude of the wavefunction. The wavefunction is assumed to be normalized to unity:  
\begin{equation}
\int | \psi (\x_{1} ,\ldots ,\x_{N} ,t)|^{2} d\x_{1} \ldots d \x_{N} = 1.
\end{equation}

In a discrete physics both position space and configuration space will be divided into cells of small finite size.  Thus we will consider the effects of dividing configuration space into cells of small finite size.  Consider dividing configuration space into cells of volume $a_{1}^{3}$\ldots $a_{N}^{3}$, where $a_{i}$ is the length characteristic of the $i$th particle represented in the wavefunction.  Here the possibility has been left open that the length of the sides of the cell in configuration space may depend on the mass or energy of the particle they correspond to.  If the length turns out to be independent of the particle and its energy then the volume of a single cell will be $a^{3N}$.  I favour the option that $a_{k}$, the length of the three sides of the cell corresponding to a single particle $k$, is proportional to the mean de Broglie wavelength of that particle.  This is because the mean de Broglie wavelength of that particle roughly characterizes the rate of change of the wavefunction with distance in the three directions in configuration space corresponding to that particle.  Further discussion of this point is beyond the scope of this paper. For more details, see \citet{leckey1998}.  

We can define a spatially discrete wavefunction that has one value per cell in configuration space: if we begin with a continuous wavefunction, this value will be the average value of the continuous wavefunction for that cell.  Now define a wavefunction that has discrete wavefunction values as well as being spatially discrete.  Take $f$ and $\theta$ as discrete:  
\begin{equation}
f(\x_{1} , \ldots , \x_{N} ,t) = n_{f\, } (\x_{1} ,\ldots , \x_{N} ,t)f_{0}\, ,     \theta (\x_{1} ,\ldots ,\x_{N} ,t) = n_{\theta} (\x_{1} ,\ldots ,\x_{N} ,t)\theta _{0},
\end{equation}
where $n_{f}$ and $n_{\theta}$ are natural numbers greater than or equal to zero, $f_{0}$ is the \textit{base magnitude} of the discrete wavefunction and $\theta_{0}$ is the \textit{base} \textit{phase} \textit{angle}. I will suppose that when the magnitude of the corresponding continuous-valued (but spatially discrete) wavefunction for a single cell, $|\psi|$, falls below $f_{0}$, the discrete wavefunction has magnitude 0.  In general, the discrete-valued wavefunction has value $nf_{0}$ where the continuous-valued wavefunction takes values between $nf_{0}$ and $(n+1)f_{0}$.

Define the \textit{relative volume} of the wavefunction as the $3N$-dimensional volume of the wavefunction in configuration space divided by the corresponding volume of a single cell in configuration space:
\begin{equation}
v=\frac{\int_{V} d{{\rm \x}}_{1} \ldots  d {{\rm \x}}_{N} }   {\int_{0}^{a_{1}} \cdots \int_{0}^{a_{N} } d\x_{1} \ldots  d{{\rm \x}}_{N} } =\frac{V}{a_{1}^{3} \ldots a_{N}^{3} }.
\end{equation}
In terms of the discrete wavefunction, $V$ is the volume in configuration space for which the wavefunction is non-zero, and the relative volume $v$ is the number of cells in configuration space occupied by the discrete wavefunction.\footnote{Here and subsequently when integrals are used, it is to be understood that these integrals apply in the case of a continuous wavefunction only.  Naturally in the case of a fully discrete physics the integrals would be replaced by sums.  Since I maintain that CCQM can stand independently of whether nature in continuous or discrete I will for simplicity use integrals, taking it as read that a sum should be substituted in the discrete case.}  The cells are of volume $a_{1}^{3}$\ldots $a_{N}^{3}$, where $a_{i}$ is the length characteristic of the $i$th particle represented in the wavefunction.

The full wavefunction of $N$ spin-half particles will be given by multiplying the spatial wavefunction by a spin vector of $2^{N}$ components, so that the total wavefunction is given by $v \times 2^{N}$ complex values.  I suggest that the relative volume provides one measure of the \textit{complexity} of the wavefunction.  Although many different measures of complexity have been proposed, in terms of computational measures of complexity it is not difficult to see why the relative volume should be a measure of complexity. The relative volume is closely connected to the amount of information required to represent the wavefunction, assuming that the amount of information required is proportional to the number of complex values used to represent the wavefunction in the position representation.  A fuller discussion of the relation between relative volume and complexity, and of the relation between relative volume and entropy, is beyond the scope of this paper---see \citet{leckey1998}.

\section{Wavefunction Collapse}
\label{Sec:WfnColl}
For reasons that will become clear, I will suppose that the volume of a wavefunction will be constricted when its relative volume reaches a certain critical value.  Thus I will suppose that there exists a \textit{critical} \textit{relative} \textit{volume} $v_{{\rm c}}$ for any wavefunction and that a wavefunction that reaches this volume will undergo a non-linear collapse, resulting in the relative volume reducing below the critical volume, or possibly remaining at the critical volume.

The existence of this critical volume may be clearly motivated in the case where the wavefunction is in reality discrete valued.   The larger the relative volume of the wavefunction in configuration space, the weaker the average magnitude of the wavefunction.  The relative volume of the wavefunction in configuration space determines the average wavefunction magnitude because the integral over configuration space of the magnitude of the wavefunction squared must be equal to unity at all times.  Thus if a wavefunction represented the state of a large number of particles, some of the particles having probability distributions spreading out in position space with time, then as the relative volume increases with time eventually the wavefunction magnitude would fall everywhere below $f_{0}$ and hence the discrete wavefunction would become zero everywhere in configuration space, which means that the system would effectively cease to exist!\footnote{The discrete wavefunction will eventually become zero everywhere if the corresponding continuous wavefunction is normalized to unity and the value of the discrete wavefunction is zero when the value of the continuous wavefunction falls below $f_{0}$.  If it were assumed instead that the discrete wavefunction were normalized to unity at all times, then the discrete wavefunction would eventually fall to zero ``nearly everywhere'', and the non-zero portions of the wavefunction would not necessarily be contiguous. Either way the result is unsatisfactory if the aim is to provide an empirically adequate theory.}

\citet[p.537]{minsky1982} discusses the problem of how to represent spherically propagating waves in position space in a discrete physics, and recognises that the tendency of the magnitude to fall to zero represents a problem for a discrete physics, but he does not suggest a solution to this problem.  Here it can be seen that a similar problem arises for discrete physics in the case of a wavefunction spreading in a $3N$ dimensional configuration space.

One way to avoid this problem would be to suppose that the base magnitude is so small that the problem would never arise - the maximum magnitude of the wavefunction would never approach $f_{0}$, even if the wavefunction covered the whole universe and represented every particle in it.  This solution could only be successful if the universe were finite in size.  Furthermore, if the base magnitude were that small then it is unlikely that the form of the laws of discrete physics would need to deviate in any significant way from those of continuous physics, yet the point of applying the discrete physics heuristic is to see what modifications to the laws of physics might arise from a discrete physics, other than simply the discreteness itself.  Thus I will consider what other way there might be of saving the phenomena, without supposing that the base magnitudes of the discrete physics are not so small that no significant conflict with the laws of continuous physics, as we currently know them, arises.

I suggest that a natural solution to the problem of spreading wavefunctions in discrete quantum mechanics is for the wavefunction to ``collapse'' to some extent when it reaches some critical relative volume, before the wavefunction becomes too weak, instantaneously (or virtually instantaneously) reducing in volume.  Alternatively, the wavefunction could split into two or more smaller wavefunctions: this alternative will be discussed in the next section.  We have supposed that the corresponding continuous wavefunction will remain normalized to unity at all times, so that when the collapse occurs the average magnitude per cell will rise.  The wavefunction would then resume spreading until it again reached the critical volume, at which time it would collapse again.  And so on.  It seems to me that this is a natural solution to the problem, so it seems that some degree of spontaneous localization of wavefunctions is a natural consequence of a discrete physics.  This is the first important result to arise from the discrete physics heuristic.

For this solution to the problem to be viable, it must be the case that the collapse does not cause deviations from quantum mechanics that are in conflict with experiments that have already been carried out.  Consider the wavefunction of a single particle in ``free'' space.  In reality, single-particle wavefunctions may be rare or non-existent, but suppose for the moment that weakly interacting particles in space are in reality represented for much of the time by wavefunctions of only one or a small number of particles per wavefunction.  (This supposition will be discussed in the next section.)  We can use these wavefunctions to put some constraints on the size of the critical relative volume, $v_{{\rm c}}$. The critical volume $V_{{\rm c}}$ must be large in size.  We know, from interference experiments that have been conducted, that a wavefunction of a free particle spreading from a coherent source can spread out over large volumes in position space before two segments of that wavefunction are combined again at the detector, producing an interference pattern.  If the critical volume $V_{{\rm c}}$ is not very large then such wavefunctions would collapse before being combined, eliminating one of the segments of the wavefunction, which would destroy the interference, leading to conflict with experimental results.

The volume that the single-particle wavefunction reduces to after collapse must also be large.  If the wavefunction when it collapsed did reduce in size by a very large degree then this would have observable consequences.  For example, a collapse of a photon wavefunction to a small volume would cause a large spreading in the distribution of its momentum components at the point the collapse occurs, in accordance with Heisenberg's uncertainty principle, so if a photon from a distant star were to collapse to a small volume before it reached us then this would affect its spectrum of frequencies in an observable way.

It should be noted that the assumption is being made that the limitation on the relative volume will apply to photons as well as particles with mass.  There are certain difficulties associated with assigning a wavefunction to a photon, but it has been demonstrated that this can be done as long as the wavefunction is interpreted slightly differently than is usual in elementary quantum mechanics.  Two ways of defining a photon wavefunction are given by \citet{bialynicki1994} and \citet{sipe1995}.  According to these definitions, the probability interpretation and normalisation condition for the photon wavefunction differ slightly from the usual ones, but not in a way that significantly affects the arguments of this paper.  The full treatment of photons is beyond the scope of CCQM: since photons are relativistic particles, they must be treated within an extension of CCQM to relativistic quantum theory.

\section{Solution to the Measurement Problem}
\label{Sec:SolMeasProb}
The type of collapse I proposed in the last section for a few-particle wavefunction is not itself supposed to comprise a solution to the measurement problem.  I have not yet discussed localizations to small regions, such as spots forming on photographic plates, that take place when ``measurements'' occur.  These kinds of localizations will emerge for many-particle systems, which will be discussed in this section, where a solution to the measurement problem will be proposed.

Consider an $N$-particle system in which each of the particles interacts relatively strongly with every other particle in the system.  I will assume that this system is represented by a single wavefunction.  Suppose that the position probability distribution of each particle in the $N$ particle interacting system covers a volume of $V_{s}$ (relative volume of $v_{s}= V_{s}/a^{3}$) in position space.  Then the corresponding relative volume of the wavefunction in configuration space will be of the order $v_{s}^{N}$, since the wavefunction will be spread over a distance of order $V_{s}^{1/3}$ in each orthogonal direction in the $3N$ dimensional space.  Thus the relative volume will tend to increase \textit{exponentially} with $N$ as the number of particles represented by a single wavefunction increases.  (Here the simplifying assumption has been made that the length of the cell-side ($a$) is the same for each particle, which is a reasonable approximation for the purpose of working out the order of magnitude dependence of the relative volume on the number of particles represented in the wavefunction.)

Thus as $N$ grows the relative volume of the wavefunction will grow very rapidly, and will readily reach the critical relative volume $v_{c}$.  Thus for an interacting many-particle system the volume covered in \textit{position space} by \textit{each particle} will tend to remain small, since a small spread in volume of the position probability distribution for each particle in position space will contribute a large amount to the relative volume of the wavefunction in configuration space - if the position probability distribution of each particle spreads out beyond a small volume in position space, then the critical relative volume will be reached, bringing about a collapse of the wavefunction, restricting each particle to the same small volume in position space again.  Thus many-particle interacting systems will have particles within them that tend to remain localized in position space, as is observed.  Furthermore, superpositions of large numbers of interacting particles spread over significant volumes in position space will be prevented from occurring by the collapse of the wavefunction.  In this way the collapse mechanism will prevent the occurrence of macroscopically distinguishable superpositions that arise in the application of the unmodified Schr\"{o}dinger equation to a typical measurement process, where no collapse is assumed.  Instead the wavefunction will collapse in such a way that these superpositions do not arise, and there will be a determinate outcome of the measurement.  Thus I claim that the critical complexity theory of the collapse of the wavefunction can provide a solution to the measurement problem.  This is the main result to arise from the discrete physics heuristic.

\section{A New Non-Linear Dynamics of Wavefunctions}
\label{Sec:NonLinDyn}
In the above I have assumed that a particle weakly interacting with its environment will be represented by a single or a few-particle wavefunction, and that a system of particles strongly interacting with each other will be represented by a many-particle wavefunction.  This assumption represents a radical departure from standard quantum mechanics.  According to standard quantum mechanics, as long as there is a non-zero strength of interaction between two systems, then those systems will be most accurately described by a single wavefunction rather than separate wavefunctions, and since there are no truly isolated systems,  it would seem that standard quantum mechanics requires that there exist only one wavefunction representing all particles.   Thus the modified quantum mechanics of CCQM must involve further modifications to linear quantum mechanics other than simply limiting the complexity of wavefunctions by introducing wavefunction collapses which localize the wavefunctions in configuration space.  The CCQM model must also involve alterations to the dynamics of linear quantum mechanics that have the effect that the number of particles represented by a single wavefunction can remain finite and change with time.   

In this modified dynamics, a system described by a single, spreading wavefunction might be postulated to evolve as follows as it comes into contact with another wavefunction.  Suppose that the position probability distributions of the particles in the first wavefunction evolve, due to the spreading of the wavefunction, in such a way that they increasingly overlap with those of particles represented by a separate wavefunction.  Then I will suppose that there is a certain probability that these systems will combine into a single wavefunction, and that the probability that they combine is related to the strength of interaction between the particles in those systems.  This strength of interaction would be determined by the types of particles involved and, under most circumstances, by the expectation value of their separation in position space.  It is natural to suppose that the greater the strength of interaction, the greater the probability that the systems combine.    The simplest way for this to occur would be for the two wavefunctions $\psi_{1}$ and $\psi_{2}$ to be replaced by their symmetrized product wavefunction:
\begin{equation}
\psi_{12} (\x_{1} ,\ldots ,\x_{N} ,t) = S \psi_{1} (\x_{1} ,\ldots ,\x_{j} ,t) \psi_{2} (\x_{j+1} ,\ldots ,\x_{N} ,t).
\end{equation}
Here the symbol $S$ refers to a symmetrization operator, which ensures that the combined wavefunction is symmetric with respect to the exchange of identical bosons between the wavefunctions, and antisymmetric with respect to the exchange of identical fermions.\footnote{It might be objected that non-interacting systems are routinely represented by product wavefunctions in standard quantum mechanics, so that the ``change'' from separate wavefunctions to a product wavefunction represents no real change at all. After all, the product and the separate wavefunctions give rise to the same empirical predictions in standard quantum mechanics.  However, it should be remembered that what is sought is a realist interpretation of the wavefunction, and under a realist interpretation a single wavefunction in a $3N$ dimensional configuration space is quite a different entity to a pair of wavefunctions in lesser dimensional configuration spaces, even if the former is equal to a product of the latter. }  This wavefunction would continue to evolve and to combine with other wavefunctions until the critical relative volume was reached, at which time the wavefunction would collapse.

When the wavefunction collapses, I assume that there is some probability that the wavefunction splits into separate wavefunctions.   One might suppose that when the collapse occurs, each sub-system has some probability of becoming represented by a separate wavefunction, and the probability of becoming separated would be greater the more weakly the particle or particles in the sub-system interact with the particles in the rest of the system.  If it happens that the wavefunction does not break up into separate wavefunctions, then the wavefunction must collapse by localising in configuration space, so bringing the relative volume below the critical volume.  One possible method for this latter kind of collapse will be discussed in Section \ref{CollapseWaveFn}.

Due to these laws for wavefunctions combining and splitting, in regions where many particles are strongly interacting the number of particles in a single wavefunction will usually be high; in regions where there are few particles, or weakly interacting particles, the number of particles in a single wavefunction will usually be low.  This is just the result that is required in order to solve the measurement problem, since we require many particles per wavefunction for many-particle, strongly interacting systems to ensure that these particles remain localised in position space.    Where there are many particles to interact with, a wavefunction will spend little time with few particles in the wavefunction, so the probability distributions of the particles will not spread very far before collapsing.   On the other hand, systems of few number of particles will be able to spread out widely before collapsing, and to be able to enter into superpositions, as required by our observations of interference effects in experiments on relatively ``free'' systems of relatively few numbers of particles.

Consider two systems ``nearby'' each other that are currently represented by separate wavefunctions.  These systems need not be considered as non-interacting in CCQM, but their interaction must be handled in this model in a different way than via an all-encompassing wavefunction in a higher-dimensional configuration space.  Just how this might occur is beyond the scope of CCQM as it now stands.  It would be hoped that the interaction between separate wavefunctions could be handled in an extension of CCQM to quantum field theory.

Before discussing the details of a possible model for those kinds of collapses that localize the wavefunction in configuration space, rather than the wavefunction splitting into two or more smaller wavefunctions, I will discuss the spontaneous localization theory of GRW, because the proposed model of collapse has features in common with the GRW theory.

\section{The GRW and CSL Models}
\label{Sec:GRWCSL}
The general idea behind the GRW model is that every particle in the universe is subject, at random times, to approximate spatial localisations.  The effect of a localisation is to cause the wavefunction representing this particle to collapse instantaneously.  Suppose that before collapse the particle is part of a system represented by a wavefunction of $N$ particles: 
\begin{equation}
\Psi (\x_{1} ,\ldots ,\x_{N} ,t).
\end{equation}
The effect on the wavefunction is given by multiplying the wavefunction by a single-particle  ``jump factor'' $j({\rm \x}^{\prime}-{\rm \x}_{k}$), where the argument \x$_{k}$ is randomly chosen from the arguments of the wavefunction, representing the localising particle.  The wavefunction after collapse is given by
\begin{equation}
\Psi_\x (\x_{1} , \ldots , \x_{N} ,t) = {\frac{\Phi_\x (\x_{1} , \ldots ,\x_{N} ,t)}{ \parallel\Phi_\x (\x_{1} , \ldots ,\x_{N} ,t) \parallel }},
\end{equation}
where 
\begin{equation}
\Phi_\x (\x_{1} ,\ldots ,\x_{N} ,t) = j(\x -\x_{k} )\Psi (\x_{1} ,\ldots ,\x_{N} ,t).
\end{equation}
We see here that the product of the wavefunction and the jump factor is renormalised at the time of collapse by the division by its norm.  GRW suggest that jump factor $j$(\x) is a Gaussian: 
\begin{equation}
j(\x ) =  (\alpha /\pi )^{3/4}\, {\rm \exp } (-\alpha\x^{2} /2),
\end{equation}
where $\alpha$ is a new constant of nature.  The effect of multiplying by the Gaussian is to approximately localise the particle within a radius of $1/\sqrt\alpha$. The localisation radius $1/\sqrt\alpha$ is assumed to take the value $10^{-5}$ cm.   The probability density of the Gaussian being centred at point \x\ is taken to be  
\begin{equation}
P(\x ,t) =   \parallel\Phi_\x (\x_{1} ,\ldots ,\x_{N} ,t) \parallel^{2}  = \int d \x_{1} \ldots d\x_{N} |\Phi_\x (\x_{1} ,\ldots ,\x_{N} ,t)|^{2}.
\end{equation}
\label{COL.PRB}This ensures that the probability of collapse is greatest where the magnitude of the wavefunction is greatest, in close agreement with the probabilistic predictions of standard quantum mechanics.   

It is assumed that the hittings for each particle occur at random times, with mean frequency $\lambda = 10^{-16}$ sec$^{-1}$, and that between each localisation the wavefunction evolves according to the Schr\"{o}dinger equation.  It follows that a single particle will suffer a collapse every 10$^{9}$ years on average, and a macroscopic system of 10$^{23}$ particles every 10$^{-7}$ sec.    

The theory is supposed to solve the measurement problem as follows: the collapses are so rare for microscopic systems containing few particles that their effects will not be observable, whereas for macroscopic systems containing large numbers of particles, the collapses will be common, and be of a type that will prevent macroscopic systems from entering into superpositions of macroscopically separate locations.

One problem perceived with the GRW method of collapse is that when a collapse takes place the existing symmetry of the wavefunction is destroyed for wavefunctions of systems containing identical particles.  The symmetry (or antisymmetry) of the wavefunction with respect to the exchange of identical particles is a requirement of standard quantum mechanics.  The problem arises for GRW because when a collapse occurs one particle is treated differently to the others - the localisation is focussed on one particle, with only some effect on other identical particles represented in the wavefunction.\footnote{\citet{dove1995} have developed versions of the original GRW model that preserve the symmetry character of the wavefunction, but these involve modifying that model so that the wavefunction involved in the collapse represents every particle in the universe.}  

In part due to the problems with the GRW model, attention has also focussed on the continuous spontaneous localisation (CSL) models that were subsequently developed \citep{gpr1990}.   The theory does preserve the symmetry character of the wavefunction.  I will not describe this theory here, as the new theory of collapse described here most closely resembles the GRW theory, and the CSL does not differ significantly from the GRW on the issues that will be central to this paper. Similarly, I will not consider relativistic versions of the GRW model \citep{tumulka2006}.

The quantum mechanical state can be represented in many mathematically equivalent ways, the representation depending on which observables are used to define the basis states of the representation.  The theory of GRW gives a privileged place to the position observable, since the localisations occur in position space rather than momentum space or the spaces of other observables. The CCQM theory also gives a privileged place to the position representation.

\section{The Collapse of the Wavefunction: A Proposal}
\label{CollapseWaveFn}
I have proposed that when the wavefunction of a system covers a certain large relative volume $v_{c}$, then there is some probability that the wavefunction will break up into separate wavefunctions, but if this does not occur then the wavefunction must ``collapse'' by localizing in configuration space, reducing in volume to some fraction $F$ of the original volume, where $F$ is some number between zero and one.  The actual fraction is not important, as long as the volume of a ``free'' wave remains large after the collapse---the fraction cannot be too small.  I will now make a tentative suggestion about the form this collapse might take. I will adapt a simple system of collapse from the GRW model.  Call this collapse model the ``jump'' model.

I will assume that when the collapse occurs, due to reaching the critical relative volume, the original wavefunction $\psi(x, t)$ is multiplied by a ``jump factor'' $j(x^{\prime}-x)$ where $j(x)$ stands for $j(\x_{1}, \ldots , \x_{N}$), and $x^{\prime}$ is the centre of collapse in configuration space.  Like GRW, I will take the jump factor $j(x$) to be a Gaussian, but in this case a (symmetrized) Gaussian in configuration space rather than a Gaussian in position space;  $j(x$) is a symmetrized product of single-particle Gaussians: 
\begin{equation}
j(\x_{1} , \ldots , \x_{N} ) = Sj(\x_{1} )\ldots  j(\x_{N} ),
\end{equation}
where 
\begin{equation}
j(\x) =  ( \varepsilon /\pi )^{3/4}\, {\rm \exp } (- \varepsilon \x^{2} /2)
\end{equation}
and $S$ refers to a symmetrization operator, which ensures that the combined wavefunction is antisymmetric with respect to the exchange of identical fermions and symmetric with respect to the exchange of identical bosons.   The value of  $\varepsilon$ will vary in each particular case, determined by the value that would be required to reduce the relative volume of the wavefunction by fraction $F$, the form of the wavefunction before collapse and the values of the constants $v_{c}$, $F$, $a_{i}$ and $f_{0}$.  I am not considering in detail experimental constraints in this paper, so I will not attempt to put bounds on the constants on the theory.  I assume the probability distribution of the collapse centre and renormalization of the wavefunction after collapse would be determined the same way as in the GRW model, so as to closely preserve the statistics of the wavefunction.  The important point to note is that the value of  $\varepsilon$ will be much smaller for systems of a few number of particles than for systems of a large number of particles, as discussed earlier.  

Unlike the GRW model, the model of collapse in CCQM preserves the symmetry of the wavefunction.  Although the wavefunction will be antisymmetric under exchange of any two identical fermions within the wavefunction, there clearly will not any symmetry under exchange of particles represented by the wavefunction with particles not represented by the wavefunction.  This is an immediate consequence of allowing wavefunctions smaller than the wavefunction of the entire universe.  However, as pointed out by  \citet[p.570]{french1978} this will not give rise to observable consequences as long as the separate wavefunctions do not overlap in configuration space; in other words, as long as there are no regions in configuration space where both have non-zero magnitude. Conversely, if separate wavefunctions do overlap significantly, then there many be observable consequences of this lack of exchange symmetry.

An alternative would be to adopt a continuous, or at least quasi-continuous collapse process, one that retains the relative volume of the wavefunction at the critical volume $v_{c}$ despite the dynamic expansion of the wavefunction.  (If nature is in fact discrete, the collapse process cannot be fully continuous, but could approximate a continuous process, that is what I mean by the term quasi-continuous.)  This collapse process would act in a similar way to the CSL theory, designed to subject the wavefunction with a series of small ``hittings'', which would have the effect of maintaining the volume of the wavefunction at $v_{c}$ despite the dynamic expansion caused by the underlying Schr\"{o}dinger evolution of the wavefunction.  Also like the CSL theory, the collapse process would occur in such a way that the wavefunction remains one contiguous whole.  Over time the wavefunction would evolve in such a way to resemble the result of the evolution of the wavefunction under the discrete ``hits'' of the jump model.  In this there is a similarity to the CSL theory: under the CSL theory, a state will evolve over time to a similar state to which the GRW theory would bring about ``all in one hit.''  Under the jump model, as the wavefunction tends to spread, its relative volume reaches $v_{c}$, collapses to some fraction of this, then grows to $v_{c}$ before collapsing again.  Under the quasi-continuous model, the relative volume would grow to $v_{c}$ then remain at that volume until such time as the wavefunction splits. A detailed treatment of the quasi-continuous collapse model is beyond the scope of this paper.

\section{Comparison of CCQM with GRW/CSL}
\label{Sec:CompCCQMGRW}
A detailed comparison of CCQM with GRW/CSL is also beyond the scope of this paper, but I will make some brief points on comparison, on top of those mentioned in the previous section.

One feature of CCQM is that, due to the fact that the relative volume grows exponentially with the number of particles, there will be a rapid transition from the  ``quantum realm'', where the position probability distributions of particles are very free to spread in position space, to the ``classical realm'' where their capacity to spread is limited.    In GRW/CSL on the other hand, this transition is linear rather than exponential in the number of particles in a system, because the rate of localization in GRW/CSL is linear in the number of particles.   Due to this difference, the transition between the two realms could take place within the realm of microscopic systems in CCQM, whereas the transition must take place within the realm of macroscopic systems in GRW/CSL.

One criticism that can be levelled at the GRW/CSL theories is that they are \textit{ad hoc} - no motivation for the modification of linear quantum mechanics that the theories involve can be given other than to produce a solution to the measurement problem.  CCQM, on the other hand, can be independently motived, as I have argued, on the basis of a fully discrete physics.  Another feature of CCQM is that it has the potential to satisfy the heuristic ``rule of simulation'' suggested by \citet{feynman1982}, who suggests as a heuristic for discovering physical laws that we only accept laws that could be simulated on finite digital computers. This provides another motivation for the theory.  I hope to discuss this and other features of CCQM in future papers.

\section{CCQM and Realism about the Wavefunction}
\label{Sec:CCQMReal}
As noted earlier, according to standard quantum mechanics, as long as there is non-zero strength of interaction between particles they must be represented by a single configuration space wavefunction, whose arguments are the positions of all the particles.   Thus it would seem to follow from standard quantum mechanics that there can only be a single wavefunction for the whole universe, and there is nothing in GRW/CSL that alters this situation.  This is to be contrasted with the case of CCQM, which admits wavefunctions that each represent a number of particles a great deal less than the number of particles in the universe.

According to the standard interpretation of a wavefunction of $N$ particles, the quantity $\mid\psi(\x_{1},\ldots ,\x_{N}, t)\mid^{2}$ is a probability density: $\mid\psi(\x_{1},\ldots ,\x_{N}, t)\mid^{2}d\x_{1}\ldots d\x_{N}$ is the probability of finding, on simultaneous measurement of the positions of each of the $N$ particles at time $t$, particle 1 within volume $d\x_{1}$ of $\x_{1}$, particle 2 within $d\x_{2}$ of $\x_{2}$,  and so on.  This interpretation is adequate if one is only interested in predicting the results of experiments, but the question is whether the wavefunction can be given a more direct realist interpretation. CCQM allows for the wavefunction itself to be interpreted realistically as a wave (or field) in configuration space, a complex valued wave in a $3N$ dimensional configuration space. (Taking spin into account, there are $2^{N}$ complex magnitudes at each point in configuration space: a vector field in configuration space.)  

This realist interpretation of the wavefunction in CCQM is consistent with the interpretation suggested by \citet{bell1990} for the wavefunction in the GRW model.  Bell suggests that the modifications to quantum mechanics introduced by the GRW model remove the concept of measurement as a primitive and allow the magnitude of the (continuous) wavefunction squared to be interpreted as the ``density of stuff'' of which the world is made.  And this is a density in a $3N$ dimensional configuration space.  

Another quantity of interest is the distribution of each particle $k$ in three-dimensional position space, given that this particle is represented by an $N$ particle wavefunction: 
\begin{equation}
g(\x_{k} ) =  \int | \psi (\x_{1} ,\ldots ,\x_{N} )  |^{2} d\x_{1} \ldots d \x_{k-1} d \x_{k+1} \ldots  d \x_{N\, .}
\end{equation}
According to the standard probabilistic interpretation of the wavefunction $g$(\x$_{k}$) is the probability density of particle $k$ in position space: $g$(\x$_{k}$)$d$\x$_{k}$ is the probability of finding on measurement particle $k$ within volume $d$\x$_{k}$ in position space.  I suggest that as well as having this probabilistic interpretation, in CCQM the quantity $g$(\x$_{k}$) can be interpreted realistically as the density of the particle $k$ in position space: it is a ``projection'' of the wavefunction from configuration space into position space.  In the case of a discrete wavefunction, the volume in position space of a single particle $k$ can be defined as the volume in position space for which the quantity $g$(\x$_{k}$) is nonzero.  (In the discrete case, the integral in the definition of $g$(\x$_{k}$) will be replaced by a sum.)  In the case of a continuous wave function, a corresponding discrete wavefunction can be defined, as described earlier, and the volume of particle $k$ can be defined in the same way: the volume in position space for which the quantity $g$(\x$_{k}$) derived from the discrete wavefunction is nonzero.

As discussed earlier, the dynamics of CCQM gives rise, for strongly interacting systems at the macroscopic level, to systems described by separate wavefunctions, each confined to a localized region, so that $g$(\x$_{k}$) will be well localized in position space for the  particles in these systems.  This corresponds to the macroscopic world as we are aware of it.  Thus we are able to provide an account of the world compatible with our experience.

\citet{ghirardi1996} and \citet{ggb1995} reject the density of stuff interpretation of Bell, and instead adopt an interpretation for CSL that involves the average mass density in three-dimensional position space. It is not entirely clear why Bell's interpretation for the entire wavefunction is rejected, but it is easy to see why the interpretation of $g$(\x$_{k}$) as the density of a particle in position space will not give a picture of reality compatible with experience in the case of CSL.  For in that model, if we presume that there is one wavefunction for the whole universe and that this wavefunction is appropriately symmetrized under the exchange of every identical particle, \textit{then every identical particle in the universe must have exactly the same position probability distribution}.  In other words, the distribution $g$(\x$_{k}$) of each identical particle must extend over the entire universe.  This is not just because the wavefunction is continuous in the GRW/CSL model and so can never fall to zero magnitude.  Even if the wavefunction were discrete, the distribution $g$(\x$_{k}$) of each particle of type $k$ would extend over the entire universe, other than regions in which there are no $k$ particles, due to the presumed symmetry character of the wavefunction.  This is why the supporters of GRW/CSL must adopt a particle-number density or mass density interpretation of the wavefunction in order to obtain a picture of reality compatible with our experience.  It seems to me to be an advantage of CCQM that it allows for a more straightforward realist interpretation of the wavefunction than is possible in the GRW/CSL models.  

\section{Conclusion}
\label{Sec:Concl}
The outlines of a new modified form of quantum mechanics, Critical Complexity Quantum Mechanics, have been presented.  The theory embraces significant departures from orthodox quantum mechanics. The first of these is the existence of wavefunctions representing finite numbers of particles, and rules for the merging and splitting of wavefunctions that do not exist in orthodox quantum mechanics.   The second is the existence of a critical relative volume which provides for an upper limit to the volume of wavefunctions in configuration space, and leads to wavefunctions splitting or collapsing, or at least having their volume constricted.

I have shown that the existence of such a critical volume follows naturally from a fully discrete physics.  Despite being philosophically inclined toward a fully discrete physics, I argue that the theory can stand even if nature is continuous, although the motivation for the theory is stronger if nature is discrete.

An advantage of the theory is that it provides for a solution to the measurement problem, at least in outline.  The solution relies on the fact that where there are many particles that strongly interact there will be many particles per wavefunction, plus the fact that the relative volume grows exponentially with the number of particles represented.  A fuller development of the theory is required in order to provide detailed empirical predictions.

The dynamic merging, splitting and collapsing of wavefunctions inherent to CCQM is based on a realist interpretation of the wavefunction, in contrast to the more orthodox anti-realist interpretations of the wavefunction. I suggest that CCQM provides an intuitively satisfactory image of quantum reality, compatible with Bell's ``density of stuff'' interpretation of the wavefunction.  The aim of providing a realistic representation of nature I see as one of the central goals of science.  I see it as well worth the price of introducing new laws governing the merging, splitting and collapsing of wavefunctions in order to obtain a satisfactory realist image of the quantum realm.

\bibliography{Bibleograph.bib}

\end{document}